%% file: iclr2026_conference.tex
\documentclass{article} 
\usepackage{iclr2026_conference}[final]
\usepackage{times}
\usepackage{sourcesanspro}

\input{math_commands.tex}

\usepackage{hyperref}
\usepackage{url}
\usepackage{booktabs}
\usepackage{multirow}
\usepackage[table]{xcolor}
\usepackage{tipa} 
\usepackage{graphicx}
\usepackage{float}
\usepackage{bbm}
\usepackage{fontawesome5}
\usepackage{xspace}

\newcommand{\modelname}{\textbf{OpenAegis}\xspace}
\xspaceaddexceptions{\footnote}

\definecolor{tablegray}{gray}{0.92}

\title{CyberFactory: Scaling Cyber Security Capabilities with Instances from the Wild}

\author{
Jian Yang\textsuperscript{1},
Haau-Sing Li\textsuperscript{2}, 
Shawn Guo\textsuperscript{3}, 
Zixi Zhao\textsuperscript{3},
Yibo Tan\textsuperscript{3},
Jiajun Wu\textsuperscript{1}, \\ \AND
Aishan Liu\textsuperscript{1},
Zhoujun Li\textsuperscript{1},
Xianglong Liu\textsuperscript{1}\thanks{Corresponding Author.}, 
Tianyu Zheng\textsuperscript{3},
Bryan Dai\textsuperscript{3},
Chengran Yang\textsuperscript{4}\thanks{Project Leader.} \\ \AND
Weifeng Lv\textsuperscript{1}
\\[4pt]
\textsuperscript{1}Beihang University \quad
\textsuperscript{2}ELLIS \quad
\textsuperscript{3}IQuest Research \quad
\textsuperscript{4}Singapore Management University \\[3pt]
}
\def\paperversion{preprint}   

\def\versionpreprint{preprint}
\def\versionfinal{final}
\ifx\paperversion\versionfinal\iclrfinalcopy\fi
\ifx\paperversion\versionpreprint\iclrfinalcopy\fi

\begin{document}

\maketitle
\vspace{-2.0em}

{
\includegraphics[height=1em]{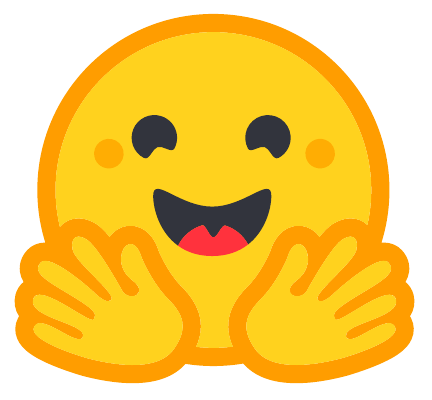}\;\ 
HuggingFace:\  \href{https://huggingface.co/collections/Multilingual-Multimodal-NLP/cyberfactory}{\textcolor[RGB]{22,59,126}{{\sffamily Multilingual-Multimodal-NLP/cyberfactory}}}
}\\
{
\includegraphics[height=1em]{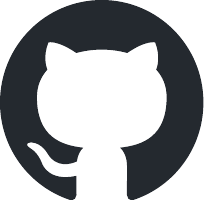}\;\ 
GitHub:\  \href{https://github.com/CSJianYang/CyberFactory}{\textcolor[RGB]{22,59,126}{{\sffamily CSJianYang/CyberFactory}}}
}

\par\vspace{0.5em}
\ifx\paperversion\versionpreprint
  \setlength{\headheight}{23pt}
  \lhead{Technical Report}
  \fancypagestyle{iquestfirstpage}{%
    \fancyhead{}%
    \lhead{\includegraphics[height=0.70cm]{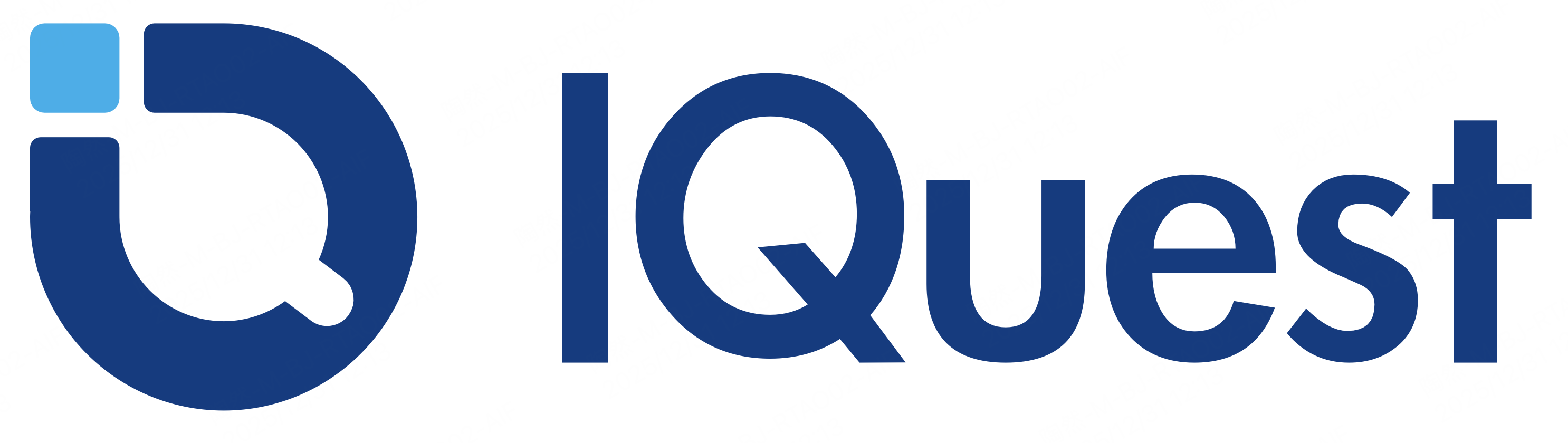}}%
  }
  \thispagestyle{iquestfirstpage}
\fi

\begin{abstract}
As large language models (LLMs) continue to advance in coding capabilities, their potential in cybersecurity has drawn increasing research attention, with closed-source LLMs (e.g., Mythos) delivering advanced cybersecurity capabilities. However, existing open-source efforts remain limited: frontier open-weight models do not provide reproducible cybersecurity training solutions, open-source training solutions focus on isolated tasks and lack scalable agentic data, and scaling agentic rollouts requires strong domain priors.
In this work, we introduce \textbf{CyberFactory}, a unified open-source framework that connects data construction, trajectory synthesis, and model training across proof-of-concept (PoC) generation, vulnerability patching, and cybersecurity question answering (CyberQA). CyberFactory transforms public vulnerability artifacts, including CVEs from the wild, into executable and verifiable task instances. It further uses a reusable vulnerability-analysis skill to guide the teacher through source inspection, problem solving with domain prior, and evidence-based validation. The resulting supervision is agentic: the model
interacts with tools and target environments and revises its solutions according to execution feedback. Using these trajectories, we train and release \modelname\footnote{\emph{Aegis} (/\textipa{"i:dZIs}/) is, in Greek mythology,
the protective shield of Zeus and Athena; the name reflects the model's
defensive, security-oriented purpose.}, which internalizes the skill-guided
procedure without requiring the skill at inference time.
On CyberGym, \modelname reaches 58.1\%
Pass@1 under a one-hour budget, improving over its Qwen~3.5 base model by 28.5
points and outperforming the evaluated general-purpose backbones
under the same scaffold.

\end{abstract}

\input{Sections/1_introduction}
\input{Sections/2_related_work}
\input{Sections/3_methodology}
\input{Sections/4_experiments}
\input{Sections/5_analysis}
\input{Sections/6_conclusion}



\bibliography{iclr2026_conference}
\bibliographystyle{iclr2026_conference}


\end{document}

%% file: math_commands.tex
\usepackage{amsmath,amsfonts,bm}

\def\eqref#1{equation~\ref{#1}}

\def\1{\bm{1}}

\DeclareMathAlphabet{\mathsfit}{\encodingdefault}{\sfdefault}{m}{sl}
\SetMathAlphabet{\mathsfit}{bold}{\encodingdefault}{\sfdefault}{bx}{n}



%% file: Sections/1_introduction.tex
\section{Introduction}

The development of large language model (LLM) has led to steady improvements in their capabilities in cybersecurity, with closed-source models delivering non-trivial performance on cybersecurity tasks~\citep{zhang2025cybench,wang2026cybergym,cybergyme2e,exploitgym}. 
Associated risks of LLMs attacking systems also arise, with emerging attacking capabilities observed under both controlled evaluations~\citep{marchand2026quantifying} and realistic deployments~\citep{openai2026huggingfaceincident}, thus calling for effective and reproducible cybersecurity methods that help inspect potential software vulnerabilities and execute defensive actions.

\begin{figure*}[t]
  \centering
  \vspace{-20pt}
  \begin{minipage}{\dimexpr\textwidth\relax}
    \centering
    \includegraphics[width=1\linewidth]{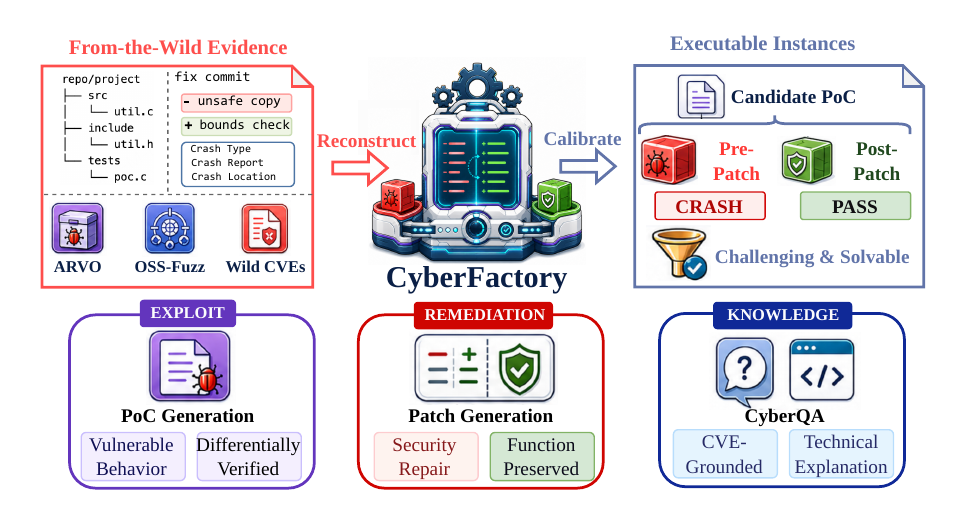}
    \vspace{-3pt}
    \caption{CyberFactory turns real-world vulnerability evidence into executable, difficulty-calibrated supervision for training \modelname on vulnerability detection, patch generation, and CyberQA.}
    \vspace{-15pt}
    \label{fig:cybergym-overview}
  \end{minipage}
\end{figure*}

Existing open-source efforts have addressed this need, yet fall short from different dimensions. Frontier open-weight models~\citep{glm5,kimik3,deepseekv4} demonstrate strong cybersecurity capabilities, but do not provide reproducible solutions for training these models. Open-source solutions span the gamut of cybersecurity capabilities like capture-the-flag (CTF), while all remain isolated efforts in disparate settings~\citep{zhou2019devign,vulnllmr,fu2022vulrepair,cyberzero,primus2025,cyberpal2024}. Moreover, in spite of the coverage of capabilities, these solutions still fail to address the need for scalable agentic data for model training. Last but not least, scaling the training of such models requires more than rollouts with simple prompts; it requires an injection of a strong domain prior as guidance~\citep{ma2026skillgen,liu2026hasp,he2026coredteam}. These gaps together motivate a key
research question: \textit{How can we build a unified and open-source approach that scales the training of cybersecurity models?}

In this paper, we introduce \textbf{CyberFactory}, a unified open-source
framework for scaling the training of cybersecurity models. Rather than
releasing model weights in isolation, CyberFactory provides an end-to-end recipe
that connects data construction, trajectory synthesis, and model training. The
framework transforms public vulnerability artifacts into executable and
verifiable instances spanning vulnerability detection, patch generation, and
CyberQA, bringing otherwise disparate tasks into a common training pipeline.
Crucially, its supervision is agentic: the teacher inspects code, invokes tools,
interacts with target environments, and revises its solutions according to
execution feedback. CyberFactory also combines crafted instance pipeline
with a reusable vulnerability-analysis skill that encodes task-independent
procedures for source inspection, problem solving with domain prior, and
evidence-based validation. Finally, We use
synthesized data to train \modelname, which internalizes the prior-guided
procedure without requiring the skill at inference time, while still demonstrating strong cybersecurity capabilities.

Under the one-hour CyberGym budget, \modelname achieves 58.1\%
Pass@1, improving over its Qwen~3.5 base model by 28.5 points and
surpassing the substantially larger GLM~5.2 and Kimi~K2.7 baselines under the
same evaluation scaffold. The benefit extends beyond the final model. When
applied to the teacher, the vulnerability-analysis skill raises GLM~5.2 Pass@1
from 43.3\% to 46.5\% despite reducing each attempt from 60 to 15 minutes,
increasing the throughput of successful training trajectories. Trajectory
analysis further shows that \modelname reproduces the skill-induced,
prior-guided workflow without receiving the skill at inference time. Together,
these results indicate that CyberFactory scales cybersecurity capability by
turning a reusable domain procedure into verifiable agentic supervision that is
subsequently internalized by the trained model.

Our contributions are as follows:
\begin{itemize}
\item We present \textbf{CyberFactory}, a unified, open-source recipe for
training cybersecurity models. It connects data construction, trajectory
synthesis, and model training across vulnerability detection, patch generation,
and CyberQA, making both the resulting supervision and its construction
procedure reproducible.
\item We develop a \textbf{scalable agentic-data pipeline} that converts CVEs
from the wild into executable, verifiable task instances and uses a reusable
vulnerability-analysis skill to guide tool-interactive trajectory synthesis.
This couples real-world vulnerability evidence with domain priors and execution
feedback rather than relying on unconstrained model rollouts.
\item We release \modelname\footnote{We release synthesized instances for reproducibility. \modelname contain risks and users should be verified on intended usage.}, a multi-task cybersecurity model trained
on the resulting agentic trajectories. Our evaluations and trajectory analyses
show that the model internalizes the skill-induced workflow without requiring
the skill at inference time, while outperforming substantially larger baselines
under the same evaluation scaffold.
\end{itemize}

%% file: Sections/2_related_work.tex
\section{Related Work}

\paragraph{Benchmarking cybersecurity capabilities of LLMs.}
\citet{bhatt2024cyberseceval} evaluate cybersecurity knowledge,
secure-code generation, and offensive capabilities, while a growing body of
work moves toward capture-the-flag (CTF) challenges and executable tasks
against real software. \citet{zhang2025cybench} curate 40 professional
CTF tasks from four competitions, add subtask decompositions for
finer-grained scoring, and show that even strong agents built on GPT-4o and
Claude~3.5~Sonnet solve only tasks that human teams complete
quickly, while the hardest challenges remain unsolved.
\citet{shao2025nyuctf} scale this format into an open, automated
framework with tool-calling over a diverse challenge database. Although
these benchmarks provide well-scoped and executable tasks, their
competition-derived challenges limit how directly they measure
real-world security work. \citet{wang2026cybergym} close this gap by
tasking agents with reproducing 1{,}507 real vulnerabilities across 188
projects from only a description and the codebase.

Complementary benchmarks broaden the executable setting.
\citet{zhu2025cvebench} evaluate agents against real web-application
vulnerabilities in reproducible environments. \citet{cybergyme2e} extend
evaluation across the full lifecycle of vulnerability discovery, PoC
generation, and patch generation, while \citet{exploitgym} test whether
agents turn triggering inputs into working exploits.

\paragraph{Reproducible vulnerability datasets.}
\citet{fan2020bigvul} link
CVE summaries to vulnerable code and corresponding code changes, while
\citet{bhandari2021cvefixes} automatically collect vulnerabilities and
their fixes from open-source repositories. \citet{hazimeh2020magma}
instrument known bugs with ground-truth reachability and triggering oracles
for fuzzer evaluation. \citet{mei2026arvo} bring reproducibility to
OSS-Fuzz by recovering over 6{,}100 real vulnerabilities, showing that reproducibility also enables
automatic patch localization and post-fix analysis.

\paragraph{LLMs for real-world software tasks.}
\citet{jimenez2024swebench} test whether
models resolve GitHub issues by editing across multiple files. \citet{yao2023react} establish
the general pattern of interleaving reasoning and environment actions.
\citet{yang2024sweagent} show how agent-computer interfaces shape
repository-level behavior, while \citet{wang2025openhands} provide an
open platform in which software agents interact with executable tool
environments. \citet{pan2025swegym} further demonstrate that executable
tasks and verifier feedback support the training of open
coding models.

\paragraph{Vulnerability repair and cybersecurity question answering.}
Neural vulnerability-repair systems learn mappings from vulnerable
functions to security patches, with \citet{fu2022vulrepair}
training a T5-based model on real-world vulnerability fixes. Cybersecurity
knowledge is typically assessed separately: \citet{liu2023secqa} introduces
a concise question-answering dataset for computer security, while
\citet{tihanyi2024cybermetric} construct multiple-choice benchmarks for
evaluating cybersecurity knowledge. CyberFactory connects these
previously separate forms of supervision with executable PoC construction in a
single multi-task data pipeline.

%% file: Sections/3_methodology.tex
\section{Methodology}

\begin{figure*}[t]
  \centering
  \includegraphics[width=\textwidth]{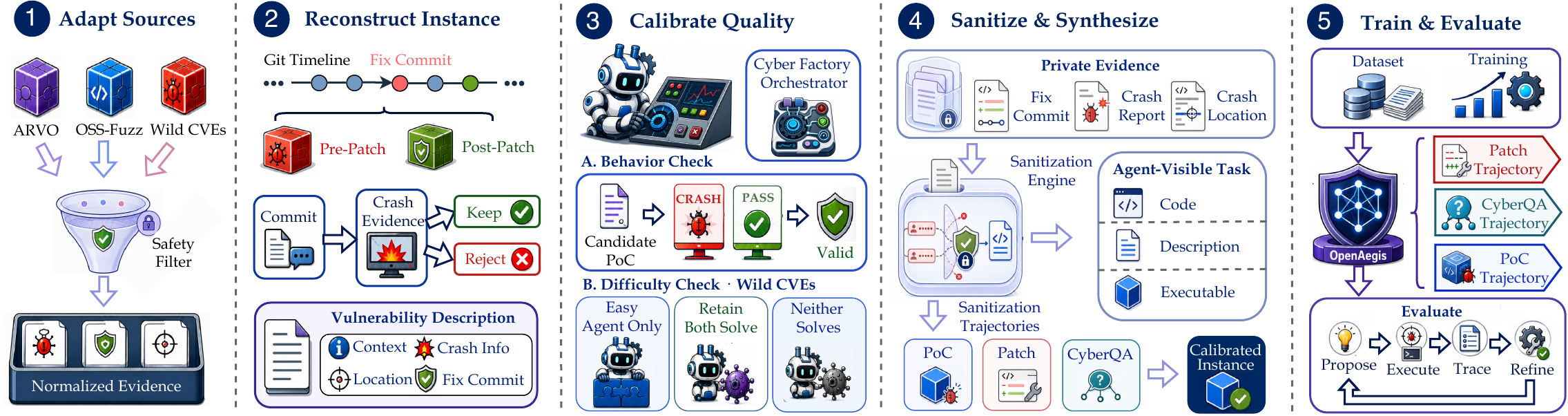}
  \vspace{-20pt}
  \caption{Cyber Factory pipeline from source adaptation to multi-task training and evaluation.}
  \label{fig:cybergym-framework}
\end{figure*}

In this section, we introduce how we synthesize data to support \modelname training. Our data synthesis pipelines span across vulnerability detection, vulnerability fixing, and question-answering of cyber security. 
~\autoref{fig:cybergym-framework} summarizes the end-to-end workflow.

\subsection{Creating Instances for PoC Construction}

A proof-of-concept (PoC) provides a concrete input that triggers a target
vulnerability. Following ground-truth benchmarks like CyberGym, we
validate a PoC differentially: it must trigger the target behavior in the
pre-patch build and not in the post-patch build~\citep{hazimeh2020magma,wang2026cybergym}. Creating instances requires building such executable environments and create a corresponding description with respect to the vulnerability.

\paragraph{Locating Vulnerabilities.}
Our PoC construction instances come from three major sources, i.e. ARVO~\citep{mei2026arvo}, OSS-Fuzz~\citep{ding2021ossfuzz}, and from-the-wild CVEs. Each source comes with a different level of difficulty of instance creation, with ARVO comes with the lowest level of difficulty and CVE-from-the-wild being the most difficult ones. Concretely, we can directly obtain vulnerabilities, i.e. a pair of pre-patch and post-patch Docker images verifiable with a groundtruth PoC. For vulnerabilities from OSS-Fuzz where only the commit that introduces the vulnerability is provided, we apply the same binary search method as mentioned in  ARVO~\citep{mei2026arvo} to locate the corresponding vulnerability fix.

From-the-wild CVEs come with the greatest level of difficulty among our collected resources. Concretely, From-the-wild CVEs typically provide affected software version ranges together with vulnerability-specific metadata, such as CWE types. 
Therefore, we create instances following a three-step pipeline. We first retain CVEs with Common Weakness Enumeration (CWE) types~\citep{martin2008cwe}. We then locate fix commits from given affected software version ranges, yielding the pre-patch version as the vulnerability version of the image and the post-patch version as the fixed one. 
The last step performs instance verification, where we filter out instances that inference models can already solve by directly generating PoCs, retaining only those that remain challenging for the models.
Note that during this step, we provide additional vulnerability-specific information, such as vulnerability types for CVE instances and crash information for OSS-Fuzz instances. These signals are used only for instance verification and are discarded during data synthesis and model training.

\paragraph{Creating Descriptions.}
With high-quality instances without a description, or query as the prompt describing the vulnerability, we first classify the corresponding vulnerability fix commit message using an LLM.
We then leverage available vulnerability-specific evidence and the fix commit to create vulnerability descriptions for those whose commit messages are of low quality, while keeping high-quality commit messages as-is.

\subsection{Patch Generation and Question-Answering}
We follow the same method applied in CVE-Factory~\citep{cvefactory} to create instances for patch generation, building on established vulnerability-fix corpora from CVE records and neural repair formulations~\citep{bhandari2021cvefixes,fu2022vulrepair}.

For question-answering (QA) data, we mainly focus QA trustworthiness. We adopt an answer-first strategy: the answer must come from a trusted source, and the LLM is responsible for turning the remaining context into a question and answer rephrase. We derive trusted answers from 
\begin{itemize}
    \item execution-derived results such as crash locations and test results, 
    \item structure-derived facts such as changed functions and call-graph relations, and
    \item text-extracted answers taken directly from authoritative reports.
\end{itemize} 
We then apply automatic checks to every sample. We use an LLM-based judge to verify that the answer can be traced back to the source and matches the original value, that the answer is not leaked in the question, and that the question identifies a unique answer. Ambiguous or invalid samples are regenerated once or discarded. These checks ensure that the QA pair faithfully reflects its source, while the actual correctness of the answer comes from the source itself.

\subsection{Skill-Guided Trajectory Synthesis}
\label{sec:skill-guided-synthesis}

\paragraph{Encoding skill for vulnerability analysis.}
Instances specify a target, but tdo not determine
how to approach the task. Direct rollouts can spend much of their budget on ad hoc input construction or repeated
attempts without a consistent analysis procedure. We therefore provide a reusable vulnerability-analysis skill when synthesizing
trajectories.

The skill encodes a task-independent workflow rather than an instance-specific solution. 
It directs the model to inspect the target and its build constraints, use appropriate analysis and testing techniques to explore candidate inputs,
validate the
resulting evidence, and revise its approach when validation fails. The model
still has to identify and trigger each vulnerability through interaction with
the environment. 
We retain trajectories with final outputs satisfying task-specific verification criteria.
Our skill conditions data synthesis; it is not supplied to
\modelname at inference time. Supervised fine-tuning on the resulting
trajectories transfers the procedure into model parameters.

\paragraph{Long-horizon context compaction.}
CyberGym rollouts can approach the 256K-token context limit before producing a
PoC. We apply the same structured compaction protocol during
training-time trajectory synthesis and inference~\citep{liu2025contexttool,jiang2023llmlingua}.
When context usage reaches 90\%, the accumulated trajectory is compressed into
a continuation state that retains verified evidence, failed attempts, open
hypotheses, generated artifacts, build status, and pending actions, while
discarding redundant logs and search results. The agent then resumes with the
same task, tools, and verification oracle; compaction changes only the
representation of its working memory.

%% file: Sections/4_experiments.tex
\section{Experiments}

\subsection{Task and Evaluation Protocol}

We evaluate on CyberGym~\citep{wang2026cybergym}, which requires synthesizing a
proof-of-concept (PoC) input that reproduces a target vulnerability given only its
natural-language description and the corresponding codebase. Let $x$ denote
a candidate input, and let a vulnerability instance be the pair of a pre-patch build
$b^{-}$ and a post-patch build $b^{+}$. We define the \emph{differential oracle}
$\mathcal{V}$ as
\begin{equation}
\label{eq:oracle}
\mathcal{V}(x) 
=\mathbbm{1}\left[\, \textsc{Crash}(b^{-}, x) \,\right]
\wedge
\mathbbm{1}\left[\, \neg\,\textsc{Crash}(b^{+}, x) \,\right],
\end{equation}
where $\textsc{Crash}(b, x)$ is true iff executing build $b$ on input $x$ triggers the
target sanitizer failure. A task is solved when the agent submits an $x^{\star}$ with
$\mathcal{V}(x^{\star})=1$: the input must trigger the vulnerability on the pre-patch
build while leaving the patched build unaffected. This two-sided criterion rules out
inputs that crash for incidental reasons, making success an objective, machine-checkable
signal rather than a heuristic match. This design follows the broader principle of
differential testing~\citep{mckeeman1998differential} and ground-truth vulnerability
oracles~\citep{hazimeh2020magma}; sanitizer instrumentation provides a concrete runtime
signal for memory errors~\citep{serebryany2012asan}.
\paragraph{The oracle as a refinement signal.}
Because ~\autoref{eq:oracle} is programmatic and decidable, an agent can test
each candidate without human supervision. The verdict and sanitizer trace
provide grounded feedback for revising subsequent candidates, turning PoC
construction into an executable propose--verify--refine loop rather than
open-ended generation.

\paragraph{Why a one-hour budget.}
PoC construction may require repeated source inspection, compilation, execution,
and validation. We therefore allow one hour per task: enough for iterative
refinement on realistic codebases while keeping evaluation bounded and
reproducible. Unless otherwise noted, every model receives the same wall-clock
budget.

\begin{table}[t]
\caption{CyberGym vulnerability reproduction under a one-hour per-task budget. Pass@1 is the
fraction of tasks for which the agent submits an input $x^{\star}$ satisfying the differential
oracle $\mathcal{V}(x^{\star})=1$ (~\autoref{eq:oracle}). Higher is better; best in \textbf{bold}.}
\vspace{5pt}
\label{tab:cybergym-main}
\centering
\small
\renewcommand{\arraystretch}{1.15}
\begin{tabular}{l c c}
\toprule
\textbf{Model} & \textbf{\# Params} & \textbf{Pass@1 (\%)} \\
\midrule
Qwen~3.5           & 397B-A17B & 29.6 \\ 
Kimi~K2.7          & 1T-A32B & 51.7 \\ 
GLM~5.2            & 744B-A40B       & 43.3 \\ 
\midrule
\rowcolor{tablegray}
\textbf{OpenAegis (ours)} & 397B-A17B & \textbf{58.1} \\
\bottomrule

\end{tabular}
\end{table}

\begin{table*}[t!]
\caption{Context management results. Higher Pass@1 and lower context exhaustion rate are better.}
\vspace{5pt}
\label{tab:compaction-ablation}
\centering
\small
\renewcommand{\arraystretch}{1.08}
\begin{tabular}{l|ccc}
\toprule
\multirow{2}{*}{\textbf{Context strategy}} & \multicolumn{2}{c}{\textbf{CyberGym Pass@1 (\%)}} & \multirow{2}{*}{\textbf{Context exhaustion (\%)}} \\
\cmidrule(lr){2-3}
& \textbf{Overall} & \textbf{Long-horizon} & \\
\midrule
\multicolumn{4}{c}{\textbf{Context management strategies}} \\
\midrule
Full history & 52.1 & 40.2 & 18.7 \\
Simple truncation & 45.6 & 36.8 & 24.5 \\
\rowcolor{tablegray}
\textbf{Compact at 90\%} & \textbf{58.1} & \textbf{48.7} & \textbf{7.0} \\
\midrule
\multicolumn{4}{c}{\textbf{Compaction trigger threshold}} \\
\midrule
Compact at 80\% & 54.5 & 45.5 & 10.4 \\
\rowcolor{tablegray}
\textbf{Compact at 90\%} & \textbf{58.1} & \textbf{48.7} & \textbf{7.0} \\
Compact at 95\% & 56.8 & 47.1 & 8.2 \\
\bottomrule
\end{tabular}
\end{table*}

\subsection{Training Details}

We initialize \modelname from the Qwen~3.5-397B-A17B checkpoint and perform full-parameter
supervised fine-tuning for three epochs. We pack sequences to a maximum length of 131,072
tokens and discard examples that exceed this limit. The micro-batch size is 1 and the global
batch size is 32. We use Adam with $\beta_1=0.9$, $\beta_2=0.95$, $\epsilon=10^{-8}$,
weight decay of 0.1, and gradient clipping at 1.0. The learning rate follows a cosine schedule,
with a peak value of $2\times10^{-5}$, a minimum of $5\times10^{-7}$, and warmup over the
first 10\% of training. Training uses BF16 computation on 256 GPUs, with tensor, context,
expert, and expert-tensor parallel sizes of 8, 2, 32, and 8, respectively.

\subsection{Baselines}

We compare \modelname with three open-weight baselines. Qwen~3.5-397B-A17B
is the initialization checkpoint and therefore measures the change produced by
our training. Kimi~K2.7 and GLM~5.2 are strong general-purpose agent backbones
with larger total and active parameter counts. Every model uses the same
scaffold, tool interface, task prompt, submission logic, differential oracle,
and one-hour wall-clock budget. None receives the vulnerability-analysis skill
at evaluation time; the skill is used only to synthesize \modelname training
trajectories.

\subsection{Main Results}

~\autoref{tab:cybergym-main} reports CyberGym Pass@1 under the one-hour
per-task budget. \modelname solves 58.1\% of the tasks, the highest rate among
the evaluated models.

The comparison with Qwen~3.5 isolates the effect of specialization on the same
397B-A17B backbone. \modelname improves Pass@1 from 29.6\% to 58.1\%, an
absolute gain of 28.5 points. This is the largest gap in the table
and shows that the training trajectories contribute substantially beyond the
capability already present in the base checkpoint.

\modelname also remains competitive with substantially larger general-purpose
models. It exceeds GLM~5.2 by 14.8 points and Kimi~K2.7 by 6.4 points,
despite using fewer total and active parameters than either model. Because all
models use the same scaffold, tools, oracle, and wall-clock budget, these
differences cannot be attributed to additional inference-time guidance or a
more permissive evaluation setup.

The aggregate result does not by itself identify which part of CyberFactory
produces the gain. We therefore examine skill-guided data synthesis in
~\autoref{tab:skill-ablation} and behavioral internalization in
~\autoref{tab:strategy-dist}. Together, the main result and subsequent
analysis test both whether the model solves more tasks and whether it has
learned the intended security workflow.

%% file: Sections/5_analysis.tex
\section{Analysis}

\subsection{Long-Horizon Context Compaction}

\begin{table}[t]
\caption{Inference-time effect of the vulnerability-analysis skill on GLM~5.2.
Pass@1 is computed over individual runs.}
\vspace{5pt}
\label{tab:skill-ablation}
\centering
\small
\renewcommand{\arraystretch}{1.15}
\begin{tabular}{l c c c}
\toprule
\textbf{GLM 5.2 configuration} & \textbf{Minutes} & \textbf{Reps} & \textbf{Pass@1 (\%)} \\
\midrule
Without analysis skill & 60 & 1 & 43.3 \\
\rowcolor{tablegray}
\textbf{With analysis skill} & \textbf{15} & \textbf{5} & \textbf{46.5} \\
\bottomrule
\end{tabular}
\end{table}

\begin{table*}[t]
\caption{Behavioral effect of providing the skill to GLM~5.2 at inference
time. 
Exploration and validation measure the use of domain-guided exploration and evidence-based validation.
Coverage is
the percentage of trajectories using each stage at least once; call counts are
normalized per trajectory.}
\vspace{5pt}
\label{tab:glm-skill-behavior}
\centering
\small
\renewcommand{\arraystretch}{1.12}
\begin{tabular}{l c c}
\toprule
\textbf{Metric} & \textbf{GLM 5.2} & \textbf{GLM 5.2 + Skill} \\
\midrule
Exploration coverage (\%) & 3.78 & \textbf{99.85} \\
Exploration calls / trajectory & 0.05 & \textbf{2.06} \\
Validation coverage (\%) & 0.13 & \textbf{98.41} \\
Validation calls / trajectory & 0.001 & \textbf{2.17} \\
Operations / shell call & 5.27 & 4.67 \\
\bottomrule
\end{tabular}
\end{table*}

~\autoref{tab:compaction-ablation} reports the context management comparison. With the same scaffold and time
budget, compact execution reaches $58.1\%$ overall Pass@1, higher than full-history execution and simple truncation. The
improvement is larger for tasks requiring more than 40 tool interactions:
Pass@1 increases by $8.5$ points, while context-exhaustion failures
decrease by $11.7$ points. The threshold ablation suggests that
$\tau=0.9$ is a useful engineering trade-off: triggering at $0.8$ compacts more
often, whereas $0.95$ leaves less room for the compaction request and following
tool calls.

\subsection{Effect of the Vulnerability-Analysis Skill}

We first isolate how the proposed skill changes the teacher used for data
synthesis by applying it to GLM~5.2 at inference time. Both conditions use the
same backbone, scaffold, tool set, and task suite, but they are not
compute-matched: the run without the skill uses one 60-minute attempt per task,
whereas the skill-guided run uses five independent 15-minute attempts.
~\autoref{tab:skill-ablation} therefore reports Pass@1
with the execution budget.

Despite receiving one quarter of the run time, skill-guided GLM~5.2 obtains
a higher Pass@1, suggesting
substantially higher synthesis throughput rather than a claim of equal-compute
improvement. Because the gain accrues during data construction, it increases
the yield of training instances without introducing an inference-time
dependency into \modelname.

The trajectory shift explains this gain. As shown in
~\autoref{tab:glm-skill-behavior}, 
the skill makes domain-guided exploration and evidence-based validation the default workflow.
Commands also become slightly less compound, suggesting
that the gain comes from choosing a scalable procedure rather than packing more
work into each command.

\subsection{Trajectory Analysis: From Skill Elicitation to Internalization}

\paragraph{Skill internalization via supervised fine-tuning.}

The clearest inherited behavior is how \modelname applies the exploration-and-validation procedure.
Supplying the skill to GLM~5.2 shifts its dominant strategy from ad hoc input construction and direct attempts toward a more systematic workflow of prior-guided exploration. Note that GLM~5.2 with skill injection can be over-dependent with our provided domain prior. 

After supervised fine-tuning,
\modelname reproduces the same directional shift relative to
Qwen3.5-397B-A17B, even though neither receives the skill at inference time.
This correspondence links inference-time elicitation in the teacher to
behavioral internalization in the trained model.


\begin{figure*}[t]
  \centering
  {\footnotesize\bfseries
  \makebox[\textwidth][l]{%
    \hspace{0.30\textwidth}\makebox[0pt][c]{Skill Elicitation}%
    \hspace{0.49\textwidth}\makebox[0pt][c]{Skill Internalization}%
  }}
  \vspace{1pt}
  \includegraphics[width=\textwidth]{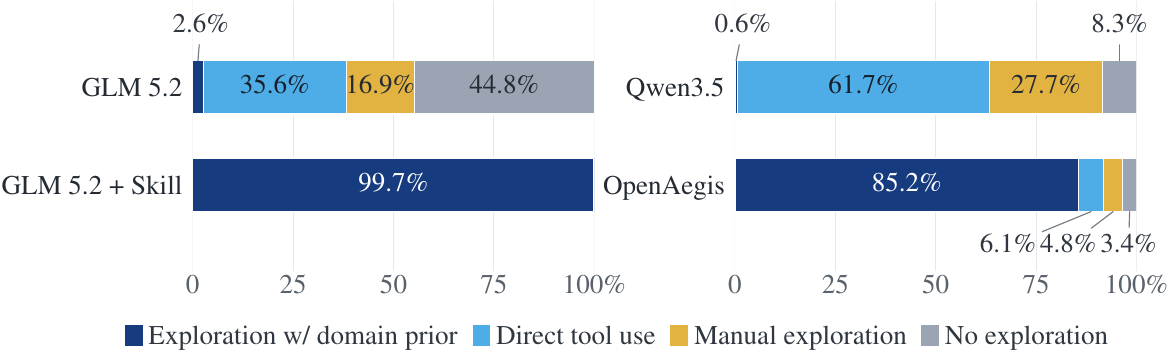}
  \caption{Distribution of exploitation strategies across explicit skill
  elicitation and training-time internalization. Each bar is normalized within
  a model; for results that do not sum to 100\%, the remaining strategies are unspecified; in-bar labels retain the reported percentages. GLM~5.2 + Skill
  receives the skill at inference time, whereas Qwen3.5 (397B-A17B) and
  \modelname do not.}
  \label{tab:strategy-dist}
\end{figure*}

\begin{table*}[t]
\caption{Use of domain-guided vulnerability analysis under explicit skill elicitation and learned internalization. Values are calls per trajectory.}
\vspace{5pt}
\label{tab:domain-prior-transfer}
\centering
\small
\renewcommand{\arraystretch}{1.12}
\begin{tabular}{l c c c c}
\toprule
\textbf{Metric} & \textbf{GLM 5.2} & \textbf{GLM 5.2 + Skill} & \textbf{Qwen3.5} & \textbf{OpenAegis} \\
\midrule
Exploration calls / trajectory & 0.05 & \textbf{2.06} & 0.01 & \textbf{1.32} \\
Validation calls / trajectory & 0.001 & \textbf{2.17} & 0.00 & \textbf{1.05} \\
\bottomrule
\end{tabular}
\end{table*}

We see the same transfer in operation frequency. As shown in
~\autoref{tab:domain-prior-transfer}, both explicit skill use and supervised
fine-tuning substantially increase exploration and validation calls. Because
\modelname does not receive the skill at inference time, this shared pattern
provides further evidence that the workflow has been internalized.

\begin{figure*}[t]
  \centering
  \includegraphics[width=0.9\textwidth]{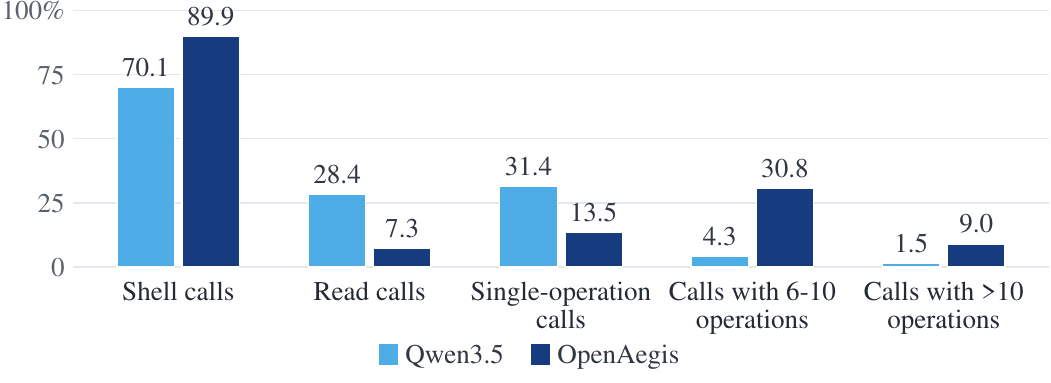}
  \caption{Tool-use and command-complexity statistics for
  Qwen3.5-397B-A17B and \modelname. Each metric is plotted independently; values
  are percentages of all tool calls.}
  \label{fig:traj-action-efficiency}
\end{figure*}

\paragraph{Action consolidation and environment-oriented tool use.}
Beyond the exploration-and-validation procedure itself, SFT changes how the model packages work into tool calls.
As shown in
~\autoref{fig:traj-action-efficiency}, \modelname shifts routine inspection from
read calls to shell calls: the former decreases from 28.4\% to 7.3\%, while the
latter rises from 70.1\% to 89.9\%. It also reduces single-operation calls from
31.4\% to 13.5\%, while increasing calls with 6--10 operations from 4.3\% to
30.8\% and calls with more than 10 operations from 1.5\% to 9.0\%. Separately,
operations per shell call increase from 2.7 to 5.5, indicating that the model
performs more environment interaction before yielding control. This contrasts
with skill-guided GLM~5.2, where operations per shell call slightly decrease
from 5.27 to 4.67. We therefore interpret action consolidation as a broader
learned behavior that emerges from the synthesized trajectories rather than a
direct fingerprint of the skill.

\paragraph{Instrumentation, verification, and submission discipline.}
Training also induces a more evidence-driven workflow. As shown in
~\autoref{tab:traj-verification}, \modelname compiles with
AddressSanitizer~\citep{serebryany2012asan} far more often and checks sanitizer
output more frequently. Submission behavior is also more selective: a larger share of \modelname
trajectories submits exactly once, while the share submitting five or more
candidates drops sharply. The previously observed rise
in prior-guided validation mirrors the workflow elicited by the skill, whereas
the changes in instrumentation and submission reflect broader behaviors learned
from successful trajectories.

\begin{table*}[t]
\caption{Instrumentation, verification, and submission behavior. Event counts
are totals over all analyzed trajectories; submission percentages report
trajectory-level shares.}
\vspace{5pt}
\label{tab:traj-verification}
\centering
\small
\renewcommand{\arraystretch}{1.12}
\begin{tabular}{l c c}
\toprule
\textbf{Metric} & \textbf{Qwen3.5} & \textbf{OpenAegis} \\
\midrule
ASAN compilation events & 155 & \textbf{1,795} \\
ASAN-output checks & 1,281 & \textbf{2,099} \\
Exactly one submission (\%) & 37.9 & \textbf{48.2} \\
At least five submissions (\%) & 10.4 & \textbf{2.0} \\
\bottomrule
\end{tabular}
\end{table*}

Taken together, the trajectory evidence establishes a continuous behavioral
link: the skill elicits a scalable, domain-guided vulnerability-analysis workflow from the teacher;
supervised fine-tuning transfers that workflow to \modelname without exposing the
skill at inference time; and the resulting model further exhibits more
consolidated tool use, stronger instrumentation, and more selective submission.
Thus, the skill serves as a data-generation mechanism whose core procedure is
internalized in model parameters rather than retained as an inference-time
dependency.

%% file: Sections/6_conclusion.tex
\section{Conclusion}

We introduced \textbf{CyberFactory}, a reproducible approach for scaling the
cybersecurity capabilities of open-weight language models by transforming
fragmented, real-world CVE artifacts into executable and verifiable supervision.
The resulting data span proof-of-concept generation, vulnerability patching, and
cybersecurity question answering, and are used to train \modelname.
CyberFactory also uses a vulnerability-analysis skill to guide the teacher
through a task-independent procedure based on source inspection, domain-guided exploration, evidence-based validation, and iterative refinement. Successful trajectories remain
grounded by executable oracles, and the skill is not needed by \modelname at
inference time.

Under a one-hour CyberGym budget, \modelname reaches 58.1\% Pass@1. This is a
28.5 point improvement over its Qwen~3.5 base model, 14.8 points above GLM~5.2,
and 6.4 points above Kimi~K2.7 under the same scaffold. The trajectory analysis
connects this aggregate gain to a change in behavior: explicit skill use moves
the teacher toward a prior-guided workflow, and \modelname exhibits the same
directional shift after training without receiving the skill at inference time.
It also uses instrumentation and validation more systematically and requires
fewer tool calls. These results indicate that verifiable trajectories can teach
both task outcomes and coherent security-analysis procedures.

CyberFactory is a step toward transparent and reproducible cybersecurity
capability development rather than a complete solution. The current evaluation
is bounded by the available CVE artifacts, benchmark coverage, and fixed
one-hour execution budget, and some targets still benefit from manual input
construction rather than a fuzzing-first strategy. Future work should broaden
the vulnerability, project, and language coverage; complete equally rigorous
evaluation of patch generation and CyberQA; and develop adaptive agents that
select among complementary analysis strategies. We hope that releasing the
pipeline and \modelname will support controlled study of useful defensive
capabilities while making their provenance and limitations easier to audit.

%% file: iclr2026_conference.bbl
\begin{thebibliography}{39}
\providecommand{\natexlab}[1]{#1}
\providecommand{\url}[1]{\texttt{#1}}
\expandafter\ifx\csname urlstyle\endcsname\relax
  \providecommand{\doi}[1]{doi: #1}\else
  \providecommand{\doi}{doi: \begingroup \urlstyle{rm}\Url}\fi

\bibitem[Bhandari et~al.(2021)Bhandari, Naseer, and Moonen]{bhandari2021cvefixes}
Guru~Prasad Bhandari, Amara Naseer, and Leon Moonen.
\newblock {CVEfixes}: Automated collection of vulnerabilities and their fixes from open-source software.
\newblock In \emph{Proceedings of the 17th International Conference on Predictive Models and Data Analytics in Software Engineering}, pp.\  30--39, 2021.
\newblock \doi{10.1145/3475960.3475985}.
\newblock URL \url{https://doi.org/10.1145/3475960.3475985}.

\bibitem[Bhatt et~al.(2024)Bhatt, Chennabasappa, Li, Nikolaidis, Song, Wan, Ahmad, Aschermann, Chen, Kapil, Molnar, Whitman, and Saxe]{bhatt2024cyberseceval}
Manish Bhatt, Sahana Chennabasappa, Yue Li, Cyrus Nikolaidis, Daniel Song, Shengye Wan, Faizan Ahmad, Cornelius Aschermann, Yaohui Chen, Dhaval Kapil, David Molnar, Spencer Whitman, and Joshua Saxe.
\newblock {CyberSecEval 2}: A wide-ranging cybersecurity evaluation suite for large language models, 2024.
\newblock URL \url{https://arxiv.org/abs/2404.13161}.

\bibitem[DeepSeek-AI et~al.(2026)DeepSeek-AI, Xu, Lin, Xue, Wang, Xu, Wu, Zhang, Lin, Dong, Ling, Lu, Zhao, Deng, Hou, Xu, Shao, Ruan, Sun, Dai, Guo, Yang, Chen, Li, Ji, Li, Wei, Lin, Yuan, Xia, Dai, Hao, Chen, Cao, Meng, Li, Yu, Zhang, Xu, Li, Liang, Zhang, Luo, Wei, Yuan, Zhang, Luo, Chen, Ji, Zhang, Ding, Tang, Cao, Gao, Qu, Zeng, Yang, Zhu, Luo, Song, Yu, Huang, Cai, Liang, Zhou, Ye, Li, Xu, Hu, Yang, Chen, Yan, Chen, Zhou, Xiang, Yuan, Cheng, Zhou, Zhu, Yu, Sun, Ran, Jiang, Qiu, Li, Zheng, Song, Dong, Gao, Guan, Zhou, Huang, Yu, Wang, Zhang, Wang, Xia, Zhang, Zhao, Guo, Luo, Ma, Zhu, Wang, Cai, Zhang, Chen, Di, Xu, Mei, Wang, Zhang, Zhang, Tang, Li, Zhou, Han, Wang, Huang, Wang, Cong, Wang, Zhang, Wang, Zhu, Li, Chen, Du, Jiang, Tian, Xu, Lu, Xu, Ge, Zhang, Pan, Wang, Chen, Yin, Xu, Shen, Zhang, Chen, Liu, Lu, Sun, Zhou, Chen, Cai, Nie, Wu, Chen, Hu, Liu, Hu, Ma, Wang, Yu, Zhou, Pan, Yu, Zhou, Ni, Yun, Jin, Pei, Ye, Lin, Ji, Cui, Yue, Yu, Wang, Zhang, Xiao, Zeng, An, Zhao, Liu, Liang, Pang, Luo, Yao,
  Gao, Yang, Huang, Hou, Zhang, Ma, Gao, He, Wang, Wang, Bi, Liu, Wang, Chen, Zhang, Nie, Sun, Wang, Cheng, Liu, Xie, Liu, Liu, Yu, Li, Yang, Zhang, Chen, Wang, Su, Chen, Lin, Fu, Yan, Wang, Ma, Luo, Zhang, Xu, Ma, Huang, Li, Li, Xu, Zhao, Sun, Wang, Qian, Shao, Yu, Zhang, Ding, Shi, Wu, Xiong, Ma, He, Tang, Zhou, Luo, Zhong, Piao, Wang, Zhang, Chen, Tan, Wei, Ma, Liu, Yang, Guo, Wu, Wu, Li, Cheng, Ou, Xu, Li, Wang, Yang, Xu, Wu, Meng, Zou, Zha, Xiong, Chen, Lin, Cao, Wang, Zhang, Yan, Lin, Gu, Luo, You, Liu, Zhou, Zhou, Huang, Wu, Wang, Zhao, Ren, Zhang, Sha, Fu, Ju, Xu, Xie, Zhang, Gao, Hao, Gou, Ma, Yan, Shao, Huang, Chen, Wu, Ren, Wu, Li, Zhang, Xu, Wang, Qu, Gu, Zhu, Li, Zhang, Xie, Gao, Wan, Pan, and Yao]{deepseekv4}
DeepSeek-AI, Anyi Xu, Bangcai Lin, Bing Xue, Bingxuan Wang, Bingzheng Xu, Bochao Wu, Bowei Zhang, Chaofan Lin, Chen Dong, Chenchen Ling, Chengda Lu, Chenggang Zhao, Chengqi Deng, Chengyu Hou, Chenhao Xu, Chenze Shao, Chong Ruan, Conner Sun, Damai Dai, Daya Guo, Dejian Yang, Deli Chen, Donghao Li, Dongjie Ji, Erhang Li, Fang Wei, Fangyun Lin, Fangzhou Yuan, Feiyu Xia, Fucong Dai, Guangbo Hao, Guanting Chen, Guoai Cao, Guolai Meng, Guowei Li, Han Yu, Han Zhang, Hanwei Xu, Hao Li, Haofen Liang, Haoling Zhang, Haoming Luo, Haoran Wei, Haotian Yuan, Haowei Zhang, Haowen Luo, Haoyu Chen, Haozhe Ji, Hengqing Zhang, Honghui Ding, Hongxuan Tang, Huanqi Cao, Huazuo Gao, Hui Qu, Hui Zeng, J~Yang, JQ~Zhu, Jia Luo, Jia Song, Jia Yu, Jialiang Huang, Jialu Cai, Jian Liang, Jiangting Zhou, Jiasheng Ye, Jiashi Li, Jiaxin Xu, Jiewen Hu, Jieyu Yang, Jin Chen, Jin Yan, Jingchang Chen, Jingli Zhou, Jingting Xiang, Jingyang Yuan, Jingyuan Cheng, Jingzi Zhou, Jinhua Zhu, Jiping Yu, Joseph Sun, Jun Ran, Junguang Jiang, Junjie Qiu,
  Junlong Li, Junmin Zheng, Junxiao Song, Kai Dong, Kaige Gao, Kang Guan, Kexing Zhou, Kezhao Huang, Kuai Yu, Lean Wang, Lecong Zhang, Lei Wang, Leyi Xia, Li~Zhang, Liang Zhao, Lihua Guo, Lingxiao Luo, Linwang Ma, Linyan Zhu, Litong Wang, Liyu Cai, Liyue Zhang, Longhao Chen, MS~Di, MY~Xu, Max Mei, Miaojun Wang, Mingchuan Zhang, Minghua Zhang, Minghui Tang, Mingming Li, Mingxu Zhou, Minmin Han, Ning Wang, Panpan Huang, Panpan Wang, Peixin Cong, Peiyi Wang, Peng Zhang, Qiancheng Wang, Qihao Zhu, Qingyang Li, Qinyu Chen, Qiushi Du, Qiwei Jiang, Rui Tian, Ruifan Xu, Ruijie Lu, Ruiling Xu, Ruiqi Ge, Ruisong Zhang, Ruizhe Pan, Runji Wang, Runqian Chen, Runqiu Yin, Runxin Xu, Ruomeng Shen, Ruoyu Zhang, Ruyi Chen, SH~Liu, Shanghao Lu, Shangmian Sun, Shangyan Zhou, Shanhuang Chen, Shaofei Cai, Shaoheng Nie, Shaoqing Wu, Shaoyuan Chen, Shengding Hu, Shengyu Liu, Shiqiang Hu, Shirong Ma, Shiyu Wang, Shuiping Yu, Shunfeng Zhou, Shuting Pan, Shuying Yu, Songyang Zhou, Tao Ni, Tao Yun, Tian Jin, Tian Pei, Tian Ye, Tianle
  Lin, Tianran Ji, Tianyi Cui, Tianyuan Yue, Tingting Yu, Tun Wang, W~Zhang, WL~Xiao, Wangding Zeng, Wei An, Weilin Zhao, Wen Liu, Wenfeng Liang, Wenjie Pang, Wenjing Luo, Wenjing Yao, Wenjun Gao, Wenkai Yang, Wenlve Huang, Wenqing Hou, Wentao Zhang, Wenting Ma, Xi~Gao, Xiang He, Xiangwen Wang, Xianzu Wang, Xiao Bi, Xiaodong Liu, Xiaohan Wang, Xiaokang Chen, Xiaokang Zhang, Xiaotao Nie, Xiaowen Sun, Xiaoxiang Wang, Xin Cheng, Xin Liu, Xin Xie, Xingchao Liu, Xingchen Liu, Xingkai Yu, Xingyou Li, Xinyu Yang, Xinyu Zhang, Xu~Chen, Xuanyu Wang, Xuecheng Su, Xueyin Chen, Xuheng Lin, Xuwei Fu, YC~Yan, YQ~Wang, YW~Ma, Yanfeng Luo, Yang Zhang, Yanhong Xu, Yanru Ma, Yanwen Huang, Yao Li, Yao Li, Yao Xu, Yao Zhao, Yaofeng Sun, Yaohui Wang, Yi~Qian, Yi~Shao, Yi~Yu, Yichao Zhang, Yifan Ding, Yifan Shi, Yijia Wu, Yiliang Xiong, Yiling Ma, Ying He, Ying Tang, Ying Zhou, Yingjia Luo, Yinmin Zhong, Yishi Piao, Yisong Wang, Yixiang Zhang, Yixiao Chen, Yixuan Tan, Yixuan Wei, Yiyang Ma, Yiyuan Liu, Yonglun Yang, Yongqiang Guo,
  Yongtong Wu, Yu~Wu, YuKun Li, Yuan Cheng, Yuan Ou, Yuanfan Xu, Yuanhao Li, Yuduan Wang, Yuehan Yang, Yuer Xu, Yuhan Wu, Yuhao Meng, Yuheng Zou, Yukun Zha, Yunfan Xiong, Yupeng Chen, Yuping Lin, Yuqian Cao, Yuqian Wang, Yushun Zhang, Yuting Yan, Yutong Lin, Yuxian Gu, Yuxiang Luo, Yuxiang You, Yuxuan Liu, Yuxuan Zhou, Yuyang Zhou, Yuzhen Huang, ZF~Wu, Zehao Wang, Zehua Zhao, Zehui Ren, Zekai Zhang, Zhangli Sha, Zhe Fu, Zhe Ju, Zhean Xu, Zhenda Xie, Zhengyan Zhang, Zheren Gao, Zhewen Hao, Zhibin Gou, Zhicheng Ma, Zhigang Yan, Zhihong Shao, Zhixian Huang, Zhixuan Chen, Zhiyu Wu, Zhizhou Ren, Zhongyu Wu, Zhuoshu Li, Zhuping Zhang, Zian Xu, Zihao Wang, Zihua Qu, Zihui Gu, Zijia Zhu, Zilin Li, Zipeng Zhang, Ziwei Xie, Ziyi Gao, Ziyi Wan, Zizheng Pan, and Zongqing Yao.
\newblock Deepseek-v4: Towards highly efficient million-token context intelligence, 2026.
\newblock URL \url{https://arxiv.org/abs/2606.19348}.

\bibitem[Ding \& Le~Goues(2021)Ding and Le~Goues]{ding2021ossfuzz}
Zhen~Yu Ding and Claire Le~Goues.
\newblock An empirical study of {OSS-Fuzz} bugs.
\newblock In \emph{2021 IEEE/ACM 18th International Conference on Mining Software Repositories}, pp.\  131--142, 2021.
\newblock \doi{10.1109/MSR52588.2021.00026}.
\newblock URL \url{https://doi.org/10.1109/MSR52588.2021.00026}.

\bibitem[Fan et~al.(2020)Fan, Li, Wang, and Nguyen]{fan2020bigvul}
Jiahao Fan, Yi~Li, Shaohua Wang, and Tien~N. Nguyen.
\newblock A {C/C++} code vulnerability dataset with code changes and {CVE} summaries.
\newblock In \emph{Proceedings of the 17th International Conference on Mining Software Repositories}, pp.\  508--512, 2020.
\newblock \doi{10.1145/3379597.3387501}.
\newblock URL \url{https://doi.org/10.1145/3379597.3387501}.

\bibitem[Fu et~al.(2022)Fu, Tantithamthavorn, Le, Nguyen, and Phung]{fu2022vulrepair}
Michael Fu, Chakkrit Tantithamthavorn, Trung Le, Van Nguyen, and Dinh Phung.
\newblock Vulrepair: A {T5}-based automated software vulnerability repair.
\newblock In \emph{Proceedings of the 30th ACM Joint European Software Engineering Conference and Symposium on the Foundations of Software Engineering}, pp.\  935--947, 2022.
\newblock \doi{10.1145/3540250.3549098}.
\newblock URL \url{https://doi.org/10.1145/3540250.3549098}.

\bibitem[GLM-5-Team et~al.(2026)GLM-5-Team, :, Zeng, Lv, Hou, Du, Zheng, Chen, Yin, Ge, Huang, Xie, Zhu, Yin, Wang, Pan, Zeng, Zhang, Wang, Chen, Zhang, Jiao, Guo, Wang, Du, Wu, Wang, Li, Fan, Zhong, Liu, Zhao, Du, Dong, Lu, Shuang-Li, Cao, Liu, Jiang, Chen, Zhang, Huang, Dong, Xu, Wei, An, Niu, Zhu, Wen, Cen, Bai, Qiao, Wang, Wang, Zhu, Liu, Li, Wang, Wen, Huang, Cai, Yu, Li, Hu, Zhang, Zhang, Lin, Yang, Wang, Ai, Zhu, Yi, Chen, Wen, Sun, Zhao, Hu, Zhang, Liu, Zhang, Peng, Tai, Zhang, Liu, Wang, Yan, Ge, Liu, Chu, Zhao, Wang, Zhao, Ren, Wang, Zhang, Gui, Zhao, Li, An, Li, Yuan, Du, Liu, Zhi, Duan, Zhou, Wei, Wang, Luo, Zhang, Sha, Xu, Wu, Ding, Chen, Li, Lin, Ta, Zou, Song, Yang, Tu, Yang, Wu, Zhang, Li, Li, Fan, Qin, Tian, Zhang, Yu, Liang, Kuang, Cheng, Li, Yan, Hu, Ling, Fan, Xia, Zhang, Zhang, Pan, Zou, Zhang, Liu, Wu, Li, Wang, Zhu, Tan, Zhou, Pan, Zhang, Su, Geng, Yan, Tan, Bi, Shen, Yang, Li, Liu, Wang, Li, Wu, Zhang, Duan, Zhang, Liu, Jiang, Yan, Zhang, Wei, Chen, Feng, Yao, Chai, Wang, Zhang, Xu,
  Huang, Wang, Li, Dong, and Tang]{glm5}
GLM-5-Team, :, Aohan Zeng, Xin Lv, Zhenyu Hou, Zhengxiao Du, Qinkai Zheng, Bin Chen, Da~Yin, Chendi Ge, Chenghua Huang, Chengxing Xie, Chenzheng Zhu, Congfeng Yin, Cunxiang Wang, Gengzheng Pan, Hao Zeng, Haoke Zhang, Haoran Wang, Huilong Chen, Jiajie Zhang, Jian Jiao, Jiaqi Guo, Jingsen Wang, Jingzhao Du, Jinzhu Wu, Kedong Wang, Lei Li, Lin Fan, Lucen Zhong, Mingdao Liu, Mingming Zhao, Pengfan Du, Qian Dong, Rui Lu, Shuang-Li, Shulin Cao, Song Liu, Ting Jiang, Xiaodong Chen, Xiaohan Zhang, Xuancheng Huang, Xuezhen Dong, Yabo Xu, Yao Wei, Yifan An, Yilin Niu, Yitong Zhu, Yuanhao Wen, Yukuo Cen, Yushi Bai, Zhongpei Qiao, Zihan Wang, Zikang Wang, Zilin Zhu, Ziqiang Liu, Zixuan Li, Bojie Wang, Bosi Wen, Can Huang, Changpeng Cai, Chao Yu, Chen Li, Chengwei Hu, Chenhui Zhang, Dan Zhang, Daoyan Lin, Dayong Yang, Di~Wang, Ding Ai, Erle Zhu, Fangzhou Yi, Feiyu Chen, Guohong Wen, Hailong Sun, Haisha Zhao, Haiyi Hu, Hanchen Zhang, Hanrui Liu, Hanyu Zhang, Hao Peng, Hao Tai, Haobo Zhang, He~Liu, Hongwei Wang, Hongxi Yan,
  Hongyu Ge, Huan Liu, Huanpeng Chu, Jia'ni Zhao, Jiachen Wang, Jiajing Zhao, Jiamin Ren, Jiapeng Wang, Jiaxin Zhang, Jiayi Gui, Jiayue Zhao, Jijie Li, Jing An, Jing Li, Jingwei Yuan, Jinhua Du, Jinxin Liu, Junkai Zhi, Junwen Duan, Kaiyue Zhou, Kangjian Wei, Ke~Wang, Keyun Luo, Laiqiang Zhang, Leigang Sha, Liang Xu, Lindong Wu, Lintao Ding, Lu~Chen, Minghao Li, Nianyi Lin, Pan Ta, Qiang Zou, Rongjun Song, Ruiqi Yang, Shangqing Tu, Shangtong Yang, Shaoxiang Wu, Shengyan Zhang, Shijie Li, Shuang Li, Shuyi Fan, Wei Qin, Wei Tian, Weining Zhang, Wenbo Yu, Wenjie Liang, Xiang Kuang, Xiangmeng Cheng, Xiangyang Li, Xiaoquan Yan, Xiaowei Hu, Xiaoying Ling, Xing Fan, Xingye Xia, Xinyuan Zhang, Xinze Zhang, Xirui Pan, Xu~Zou, Xunkai Zhang, Yadi Liu, Yandong Wu, Yanfu Li, Yidong Wang, Yifan Zhu, Yijun Tan, Yilin Zhou, Yiming Pan, Ying Zhang, Yinpei Su, Yipeng Geng, Yong Yan, Yonglin Tan, Yuean Bi, Yuhan Shen, Yuhao Yang, Yujiang Li, Yunan Liu, Yunqing Wang, Yuntao Li, Yurong Wu, Yutao Zhang, Yuxi Duan, Yuxuan Zhang,
  Zezhen Liu, Zhengtao Jiang, Zhenhe Yan, Zheyu Zhang, Zhixiang Wei, Zhuo Chen, Zhuoer Feng, Zijun Yao, Ziwei Chai, Ziyuan Wang, Zuzhou Zhang, Bin Xu, Minlie Huang, Hongning Wang, Juanzi Li, Yuxiao Dong, and Jie Tang.
\newblock Glm-5: from vibe coding to agentic engineering, 2026.
\newblock URL \url{https://arxiv.org/abs/2602.15763}.

\bibitem[Hazimeh et~al.(2020)Hazimeh, Herrera, and Payer]{hazimeh2020magma}
Ahmad Hazimeh, Adrian Herrera, and Mathias Payer.
\newblock Magma: A ground-truth fuzzing benchmark.
\newblock \emph{Proceedings of the ACM on Measurement and Analysis of Computing Systems}, 4\penalty0 (3):\penalty0 1--29, 2020.
\newblock \doi{10.1145/3428334}.
\newblock URL \url{https://doi.org/10.1145/3428334}.

\bibitem[He et~al.(2026)He, Fox, Miculicich, Friedli, Fabian, Gokturk, Tang, Lee, Pfister, and Le]{he2026coredteam}
Pengfei He, Ash Fox, Lesly Miculicich, Stefan Friedli, Daniel Fabian, Burak Gokturk, Jiliang Tang, Chen-Yu Lee, Tomas Pfister, and Long~T. Le.
\newblock Co-redteam: Orchestrated security discovery and exploitation with {LLM} agents.
\newblock In \emph{Proceedings of the 43rd International Conference on Machine Learning}, 2026.
\newblock URL \url{https://research.google/pubs/co-redteam-orchestrated-security-discovery-and-exploitation-with-llm-agents/}.

\bibitem[Jiang et~al.(2023)Jiang, Wu, Lin, Yang, and Qiu]{jiang2023llmlingua}
Huiqiang Jiang, Qianhui Wu, Chin-Yew Lin, Yuqing Yang, and Lili Qiu.
\newblock {LLMLingua}: Compressing prompts for accelerated inference of large language models.
\newblock In \emph{Proceedings of the 2023 Conference on Empirical Methods in Natural Language Processing}, pp.\  13358--13376, 2023.
\newblock \doi{10.18653/v1/2023.emnlp-main.825}.
\newblock URL \url{https://aclanthology.org/2023.emnlp-main.825/}.

\bibitem[Jimenez et~al.(2024)Jimenez, Yang, Wettig, Yao, Pei, Press, and Narasimhan]{jimenez2024swebench}
Carlos~E. Jimenez, John Yang, Alexander Wettig, Shunyu Yao, Kexin Pei, Ofir Press, and Karthik Narasimhan.
\newblock {SWE}-bench: Can language models resolve real-world github issues?
\newblock In \emph{The Twelfth International Conference on Learning Representations}, 2024.
\newblock URL \url{https://arxiv.org/abs/2310.06770}.

\bibitem[Levi et~al.(2024)Levi, Alluouche, Ohayon, and Puzanov]{cyberpal2024}
Matan Levi, Yair Alluouche, Daniel Ohayon, and Anton Puzanov.
\newblock Cyberpal.ai: Empowering {LLMs} with expert-driven cybersecurity instructions, 2024.
\newblock URL \url{https://arxiv.org/abs/2408.09304}.

\bibitem[Liu et~al.(2026)Liu, Ming, Joty, and Zhao]{liu2026hasp}
Hongjun Liu, Yifei Ming, Shafiq Joty, and Chen Zhao.
\newblock Harnessing {LLM} agents with skill programs, 2026.
\newblock URL \url{https://arxiv.org/abs/2605.17734}.

\bibitem[Liu et~al.(2025)Liu, Yang, Jiang, Li, Guo, Liu, and Dai]{liu2025contexttool}
Shukai Liu, Jian Yang, Bo~Jiang, Yizhi Li, Jinyang Guo, Xianglong Liu, and Bryan Dai.
\newblock Context as a tool: Context management for long-horizon {SWE}-agents, 2025.
\newblock URL \url{https://arxiv.org/abs/2512.22087}.

\bibitem[Liu(2023)]{liu2023secqa}
Zefang Liu.
\newblock {SecQA}: A concise question-answering dataset for evaluating large language models in computer security, 2023.
\newblock URL \url{https://arxiv.org/abs/2312.15838}.

\bibitem[Luo et~al.(2026)Luo, Zhang, Zhou, Huang, Xiao, Zhu, Ma, Yue, Yue, Zeng, and Che]{cvefactory}
Xianzhen Luo, Jingyuan Zhang, Shiqi Zhou, Jinyang Huang, Chuan Xiao, Qingfu Zhu, Zhiyuan Ma, Xing Yue, Yang Yue, Wencong Zeng, and Wanxiang Che.
\newblock Cve-factory: Scaling expert-level agentic tasks for code security vulnerability, 2026.
\newblock URL \url{https://arxiv.org/abs/2602.03012}.

\bibitem[Ma et~al.(2026)Ma, Huang, Bao, Zhuang, Shukla, Galley, Zhang, and Feuerriegel]{ma2026skillgen}
Yuchen Ma, Yue Huang, Han Bao, Haomin Zhuang, Swadheen Shukla, Michel Galley, Xiangliang Zhang, and Stefan Feuerriegel.
\newblock Skillgen: Verified inference-time agent skill synthesis, 2026.
\newblock URL \url{https://arxiv.org/abs/2605.10999}.

\bibitem[Marchand et~al.(2026)Marchand, Cathain, Wynne, Giavridis, Deverett, Wilkinson, Gwartz, and Coppock]{marchand2026quantifying}
Rahul Marchand, Art~O Cathain, Jerome Wynne, Philippos~Maximos Giavridis, Sam Deverett, John Wilkinson, Jason Gwartz, and Harry Coppock.
\newblock Quantifying frontier {LLM} capabilities for container sandbox escape.
\newblock In \emph{Forty-third International Conference on Machine Learning}, 2026.
\newblock URL \url{https://openreview.net/forum?id=19AbP986bv}.

\bibitem[Martin \& Barnum(2008)Martin and Barnum]{martin2008cwe}
Robert~A. Martin and Sean Barnum.
\newblock Common weakness enumeration ({CWE}) status update.
\newblock \emph{ACM SIGAda Ada Letters}, 28\penalty0 (1):\penalty0 88--91, 2008.
\newblock \doi{10.1145/1387830.1387835}.
\newblock URL \url{https://doi.org/10.1145/1387830.1387835}.

\bibitem[McKeeman(1998)]{mckeeman1998differential}
William~M. McKeeman.
\newblock Differential testing for software.
\newblock \emph{Digital Technical Journal}, 10\penalty0 (1):\penalty0 100--107, 1998.
\newblock URL \url{https://www.bitsavers.org/pdf/dec/dtj/dtj_v10-01_1998.pdf}.

\bibitem[Mei et~al.(2026)Mei, Castillo, Singaria, Xi, Benchikh, Bao, Wang, Shoshitaishvili, Doup{\'e}, Pearce, and Dolan-Gavitt]{mei2026arvo}
Xiang Mei, Jordi~Del Castillo, Pulkit~Singh Singaria, Haoran Xi, Abdelouahab Benchikh, Tiffany Bao, Ruoyu Wang, Yan Shoshitaishvili, Adam Doup{\'e}, Hammond Pearce, and Brendan Dolan-Gavitt.
\newblock {ARVO}: Atlas of reproducible vulnerabilities for open source software.
\newblock In \emph{IEEE European Symposium on Security and Privacy (EuroS\&P)}, 2026.
\newblock URL \url{https://arxiv.org/abs/2408.02153}.

\bibitem[Nie et~al.(2025)Nie, Li, Guo, Jiang, Wang, Li, Song, and Guo]{vulnllmr}
Yuzhou Nie, Hongwei Li, Chengquan Guo, Ruizhe Jiang, Zhun Wang, Bo~Li, Dawn Song, and Wenbo Guo.
\newblock Vulnllm-r: Specialized reasoning {LLM} with agent scaffold for vulnerability detection, 2025.
\newblock URL \url{https://arxiv.org/abs/2512.07533}.

\bibitem[{OpenAI}(2026)]{openai2026huggingfaceincident}
{OpenAI}.
\newblock Openai and hugging face partner to address security incident during model evaluation.
\newblock OpenAI Security Blog, July 2026.
\newblock URL \url{https://openai.com/index/hugging-face-model-evaluation-security-incident/}.

\bibitem[Pan et~al.(2025)Pan, Wang, Neubig, Jaitly, Ji, Suhr, and Zhang]{pan2025swegym}
Jiayi Pan, Xingyao Wang, Graham Neubig, Navdeep Jaitly, Heng Ji, Alane Suhr, and Yizhe Zhang.
\newblock Training software engineering agents and verifiers with {SWE}-gym.
\newblock In \emph{Proceedings of the 42nd International Conference on Machine Learning}, volume 267 of \emph{Proceedings of Machine Learning Research}, pp.\  47717--47737, 2025.
\newblock URL \url{https://proceedings.mlr.press/v267/pan25g.html}.

\bibitem[Serebryany et~al.(2012)Serebryany, Bruening, Potapenko, and Vyukov]{serebryany2012asan}
Konstantin Serebryany, Derek Bruening, Alexander Potapenko, and Dmitry Vyukov.
\newblock {AddressSanitizer}: A fast address sanity checker.
\newblock In \emph{2012 USENIX Annual Technical Conference}, pp.\  309--318. USENIX Association, 2012.
\newblock URL \url{https://www.usenix.org/conference/atc12/addresssanitizer-fast-address-sanity-checker}.

\bibitem[Shao et~al.(2024)Shao, Jancheska, Udeshi, Dolan-Gavitt, Xi, Milner, Chen, Yin, Garg, Krishnamurthy, Khorrami, Karri, and Shafique]{shao2025nyuctf}
Minghao Shao, Sofija Jancheska, Meet Udeshi, Brendan Dolan-Gavitt, Haoran Xi, Kimberly Milner, Boyuan Chen, Max Yin, Siddharth Garg, Prashanth Krishnamurthy, Farshad Khorrami, Ramesh Karri, and Muhammad Shafique.
\newblock {NYU} {CTF} bench: A scalable open-source benchmark dataset for evaluating {LLM}s in offensive security.
\newblock In \emph{Advances in Neural Information Processing Systems (NeurIPS)}, 2024.
\newblock URL \url{https://arxiv.org/abs/2406.05590}.

\bibitem[Shi et~al.(2026)Shi, Rheem, Jiang, Wang, Riega, Wang, Jiang, Cheung, Tai, Cha, Tu, Han, Wang, He, Guo, and Song]{cybergyme2e}
Tianneng Shi, Robin Rheem, Dongwei Jiang, Mona Wang, Francisco De~La Riega, Zhun Wang, Jingzhi Jiang, Alexander Cheung, Sean Tai, Jonah Cha, Jianhong Tu, Gabriel Han, Chenguang Wang, Jingxuan He, Wenbo Guo, and Dawn Song.
\newblock Cybergym-e2e: Scalable real-world benchmark for ai agents' end-to-end cybersecurity capabilities, 2026.
\newblock URL \url{https://arxiv.org/abs/2606.04460}.

\bibitem[Team et~al.(2026)Team, Bai, Bai, Bao, C., Cai, Cai, Cao, Cao, Chai, Charles, Che, Chen, Chen, Chen, Chen, Chen, Chen, Chen, Chen, Chen, Chen, Chen, Chen, Chen, Chen, Chen, Chen, Chen, Chen, Chen, Chen, Cheng, Cheng, Cui, Cui, Dai, Deng, Ding, Ding, Ding, Dong, Dong, Dong, Dong, Du, Du, Du, Du, Du, Fan, Feng, Feng, Feng, Fu, Fu, Gao, Gao, Gao, Gao, Gao, Geng, Gong, Gong, Gong, Gong, Gu, Gu, Guan, Guo, Guo, Guo, Guo, Hao, Hao, Hao, He, He, He, He, He, He, He, He, He, Hong, Hong, Hu, Hu, Hu, Hu, Hu, Hu, Hua, Huang, Huang, Huang, Huang, Huang, Huang, Huang, Huang, Hui, Jia, Jiang, Jiang, Jiang, Jin, Jin, Jing, Kong, Lai, Li, Li, Li, Li, Li, Li, Li, Li, Li, Li, Li, Li, Li, Li, Li, Li, Li, Li, Li, Li, Li, Li, Li, Li, Lin, Lin, Lin, Lin, Lin, Liu, Liu, Liu, Liu, Liu, Liu, Liu, Liu, Liu, Liu, Liu, Liu, Liu, Liu, Liu, Lu, Lu, Lu, Lu, Lu, Luo, Luo, Luo, Luo, Lyu, Lyu, Mao, Mei, Men, Ni, Niu, Pan, Peng, Qi, Qin, Qin, Qin, Qiu, Qiu, Qiu, Qu, Qu, Shang, Shao, Shen, Shi, Shi, Shi, Shi, Siu, Song, Song, Su, Su, Su,
  Sui, Sun, Sun, Sun, Sun, Sun, Sun, Tai, Tang, Tang, Tang, Tang, Tian, Tian, Tian, Tu, Wang, Wang, Wang, Wang, Wang, Wang, Wang, Wang, Wang, Wang, Wang, Wang, Wang, Wang, Wang, Wang, Wang, Wang, Wang, Wang, Wang, Wang, Wang, Wang, Wang, Wang, Wang, Wang, Wang, Wang, Wang, Wang, Wang, Wang, Wang, Wang, Wang, Wang, Wang, Wang, Wang, Wei, Wei, Wei, Wen, Wu, Wu, Wu, Wu, Wu, Wu, Wu, Wu, Wu, Xian, Xiang, Xiang, Xiao, Xiao, Xiao, Xie, Xie, Xie, Xie, Xing, Xiong, Xu, Xu, Xu, Xu, Xu, Xu, Xu, Xu, Xu, Xu, Xu, Xu, Xu, Xu, Xu, Xu, Xu, Xu, Xue, Yan, Yan, Yang, Yang, Yang, Yang, Yang, Yang, Yang, Yang, Yang, Yang, Yang, Yang, Yang, Yang, Yang, Yang, Yao, Ye, Ye, Ye, Ye, Yin, Yin, Yin, Yu, Yu, Yu, Yu, Yu, Yu, Yuan, Yuan, Yue, Yue, Yue, Zha, Zhan, Zhang, Zhang, Zhang, Zhang, Zhang, Zhang, Zhang, Zhang, Zhang, Zhang, Zhang, Zhang, Zhang, Zhang, Zhang, Zhang, Zhang, Zhang, Zhang, Zhang, Zhang, Zhang, Zhang, Zhang, Zhang, Zhang, Zhang, Zhang, Zhang, Zhang, Zhang, Zhao, Zhao, Zhao, Zhao, Zhao, Zhao, Zhao, Zhao, Zhao, Zhao, Zhao,
  Zheng, Zheng, Zheng, Zheng, Zheng, Zhong, Zhong, Zhong, Zhou, Zhou, Zhou, Zhou, Zhou, Zhou, Zhou, Zhu, Zhu, Zhu, Zhu, Zhu, Zhu, Zhuang, Zhuang, and Zu]{kimik3}
Kimi Team, Tongtong Bai, Yifan Bai, Yiping Bao, M.~C., Jianfeng Cai, Xinyuan Cai, Peizhou Cao, Yuxuan Cao, Ziwei Chai, Y.~Charles, H.~S. Che, Guanduo Chen, Guangyu Chen, Guanzheng Chen, Huarong Chen, Jia Chen, Jianlong Chen, Jun Chen, Kexin Chen, Peng Chen, Ruijue Chen, Wentao Chen, Xin Chen, Yang Chen, Yanru Chen, Yifei Chen, Yingjiang Chen, Yuankun Chen, Yujie Chen, Yutian Chen, Zhirong Chen, Dazhi Cheng, Yean Cheng, Jialei Cui, Jingbing Cui, Anqi Dai, Jiaqi Deng, Hao Ding, Rui Ding, Shaofeng Ding, Mengfan Dong, Mengnan Dong, Yuhao Dong, Yuxin Dong, Angang Du, Chenzhuang Du, Dikang Du, Jusen Du, Yulun Du, Yu~Fan, Jing Feng, Qiulin Feng, Yichen Feng, Kelin Fu, Qiang Fu, Fuxuan Gao, Hongcheng Gao, Jingyue Gao, Tong Gao, Weijia Gao, Shangyi Geng, Jie Gong, Linhu Gong, Shengao Gong, Xiaochen Gong, Qizheng Gu, Yicheng Gu, Shuhao Guan, Haiqing Guo, Shiqi Guo, Xiang Guo, Zhengyan Guo, Beixi Hao, Wenxin Hao, Xiaoru Hao, Dailan He, Haotian He, Lehan He, Qi~He, Weiran He, Xinran He, Xinyi He, Yibo He, Yunjia He, Chao
  Hong, Tiange Hong, Hao Hu, Jiaxi Hu, Ruikun Hu, Weiming Hu, Yangyang Hu, Zhenxing Hu, Liang Hua, Jinbin Huang, Ke~Huang, Ruiyuan Huang, Siying Huang, Weixiao Huang, Yan Huang, Zhengjie Huang, Zhiqi Huang, Yulong Hui, Chaobo Jia, Yutong Jiang, Zhejun Jiang, Zuoyou Jiang, Wenyi Jin, Xinyi Jin, Yu~Jing, Huanjun Kong, Guokun Lai, Aidi Li, Cheng Li, Chengyuan Li, Cong Li, Fang Li, Guanyu Li, Haoyang Li, Jia Li, Junxiong Li, Lei Li, Letian Li, Lincan Li, Weihong Li, Wentao Li, Xintong Li, Yang Li, Yishen Li, Yiwei Li, Yuxiao Li, Zhaowei Li, Zhaoxi Li, Zheming Li, Zhengxiao Li, Zhiyuan Li, Jiawei Lin, Xiaohan Lin, Yibo Lin, Zichao Lin, Ziyan Lin, Bill Liu, Boxiao Liu, Chuan Liu, Liang Liu, Shaowei Liu, Shudong Liu, Shuran Liu, Tianwei Liu, Weizhou Liu, Yangyang Liu, Yanming Liu, Yibo Liu, Yipeng Liu, Zhengying Liu, Zhiheng Liu, Enzhe Lu, Haoyu Lu, Linqiang Lu, Tingzhan Lu, Zhiyuan Lu, Aotian Luo, G.~Luo, Junyu Luo, Yifan Luo, B.~Lyu, Wenzhou Lyu, Shaoguang Mao, Yuan Mei, Xin Men, Minqing Ni, Yixuan Niu, Siyuan
  Pan, Shujun Peng, Zhangyang Qi, Ruoyu Qin, ZeChao Qin, Zeyu Qin, Haiquan Qiu, Jianxin Qiu, Jiezhong Qiu, Bowen Qu, Yuhao Qu, Zeyu Shang, Youbo Shao, Han Shen, Jincheng Shi, Juanfeng Shi, Lidong Shi, Shengyuan Shi, Wingchun Siu, Pengwei Song, Xiaoxi Song, Jianlin Su, Yunfeng Su, Zhaochen Su, Lin Sui, Jingsong Sun, Junyao Sun, Shaoning Sun, Shuzhe Sun, Tongyu Sun, Yujun Sun, Yunpeng Tai, Chuning Tang, Heyi Tang, Sirui Tang, Zecheng Tang, Chaoran Tian, Rongpeng Tian, Yu~Tian, Wei Tu, Chensi Wang, Chuang Wang, Chunjie Wang, Dinglu Wang, Feng Wang, Hailong Wang, Haiming Wang, Hao Wang, Hao Wang, Huaqing Wang, Hui Wang, Jiayi Wang, Jinglong Wang, Jinhong Wang, Jiuzheng Wang, Linian Wang, Shaobo Wang, Shenzhi Wang, Shuyi Wang, Si~Wang, Siyuan Wang, Tianfu Wang, Wenjue Wang, Xingran Wang, Xinmei Wang, Xinyuan Wang, Xusheng Wang, Yalin Wang, Yangkun Wang, Yao Wang, Yaoyu Wang, Yejie Wang, Yiqin Wang, Yucheng Wang, Yuzhi Wang, Zhaoji Wang, Zhaowei Wang, Zhengtao Wang, Zhenhao Wang, Zhongsheng Wang, Zifan Wang, Chu
  Wei, Ming Wei, Shouxin Wei, Zichen Wen, Fan Wu, Haoning Wu, Rucong Wu, Wenhao Wu, Xiaoxue Wu, Yingcong Wu, Yongqi Wu, Yuxin Wu, Zijian Wu, Xinglang Xian, Chenxuan Xiang, Yuye Xiang, Bocheng Xiao, Chenjun Xiao, Xin Xiao, Jin Xie, Xiaotong Xie, Yifeng Xie, Zhe Xie, Bowei Xing, Yiming Xiong, Baosheng Xu, Boyu Xu, Jiale Xu, Jianfan Xu, Jing Xu, Jinjing Xu, L.~H. Xu, Qingtao Xu, Shuyao Xu, Suting Xu, Tiantian Xu, Tianxiang Xu, Weixin Xu, Xinran Xu, Yangchuan Xu, Ye~Xu, Yueni Xu, Ziyao Xu, Haonan Xue, Junjie Yan, Yaoyao Yan, Fan Yang, Guangyao Yang, Hao Yang, Junwei Yang, Ruoyu Yang, Wenjie Yang, Xiaofei Yang, Xinyu Yang, Yi~Yang, Yiling Yang, Ying Yang, Yuchen Yang, Zhen Yang, Zhilin Yang, Zian Yang, Zuhao Yang, Haotian Yao, Dan Ye, Haoran Ye, Wenjie Ye, Zhanbo Ye, Bohong Yin, Haoxiang Yin, Xietong Yin, Chengzhen Yu, Haozhen Yu, Longhui Yu, Shengnan Yu, Shuying Yu, Tianxiang Yu, Enming Yuan, Mengjie Yuan, Tongtian Yue, Wei Yue, Yang Yue, Dunyuan Zha, Haobing Zhan, B.~H. Zhang, Dehao Zhang, Fei Zhang, Hao Zhang,
  Haoyuan Zhang, Huanyu Zhang, Jiapei Zhang, Jiaxuan Zhang, Jin Zhang, Kaiyi Zhang, Miaozhen Zhang, Puqi Zhang, Qinglei Zhang, Rong Zhang, Rui Zhang, Shaoshuai Zhang, Shiyi Zhang, Xiaobin Zhang, Xiaoyun Zhang, Y.~Zhang, Yangkun Zhang, Ye~Zhang, Yichi Zhang, Yikun Zhang, Yizhi Zhang, Yongting Zhang, Yu~Zhang, Yutao Zhang, Yutong Zhang, Zheng Zhang, Zijing Zhang, Bin Zhao, Chenguang Zhao, Feifan Zhao, Jinglun Zhao, Jinxiang Zhao, Shuai Zhao, Wenshuo Zhao, Xiangyu Zhao, Xuanle Zhao, Yikai Zhao, Zijia Zhao, Haozhi Zheng, Huabin Zheng, Ruihan Zheng, Shaojie Zheng, Tengyang Zheng, Haofeng Zhong, Lei Zhong, Longguang Zhong, M.~Zhou, Qiankang Zhou, Runjie Zhou, Ruozhang Zhou, Xinyu Zhou, Yiqiao Zhou, Zaida Zhou, Jinguo Zhu, Liya Zhu, Xinhao Zhu, Yangjunfeng Zhu, Yuxuan Zhu, Zhen Zhu, Chen Zhuang, Weiyu Zhuang, and Xinxing Zu.
\newblock Kimi k3: Open frontier intelligence, 2026.
\newblock URL \url{https://arxiv.org/abs/2607.24653}.

\bibitem[Tihanyi et~al.(2024)Tihanyi, Ferrag, Jain, Bisztray, and Debbah]{tihanyi2024cybermetric}
Norbert Tihanyi, Mohamed~Amine Ferrag, Ridhi Jain, Tamas Bisztray, and Merouane Debbah.
\newblock {CyberMetric}: A benchmark dataset based on retrieval-augmented generation for evaluating {LLM}s in cybersecurity knowledge, 2024.
\newblock URL \url{https://arxiv.org/abs/2402.07688}.

\bibitem[Wang et~al.(2025)Wang, Li, Song, Xu, Tang, Zhuge, Pan, Song, Li, Singh, Tran, Li, Ma, Zheng, Qian, Shao, Muennighoff, Zhang, Hui, Lin, Brennan, Peng, Ji, and Neubig]{wang2025openhands}
Xingyao Wang, Boxuan Li, Yufan Song, Frank~F. Xu, Xiangru Tang, Mingchen Zhuge, Jiayi Pan, Yueqi Song, Bowen Li, Jaskirat Singh, Hoang Tran, Fuqiang Li, Ren Ma, Mingzhang Zheng, Bill Qian, Daniel Shao, Niklas Muennighoff, Yizhe Zhang, Binyuan Hui, Junyang Lin, Robert Brennan, Hao Peng, Heng Ji, and Graham Neubig.
\newblock {OpenHands}: An open platform for {AI} software developers as generalist agents.
\newblock In \emph{The Thirteenth International Conference on Learning Representations}, 2025.
\newblock URL \url{https://proceedings.iclr.cc/paper_files/paper/2025/hash/a4b6ad6b48850c0c331d1259fc66a69c-Abstract-Conference.html}.

\bibitem[Wang et~al.(2026{\natexlab{a}})Wang, Schiller, Li, Narayana, Nasr, Carlini, Qi, Wallace, Bursztein, Invernizzi, Thomas, Shoshitaishvili, Guo, He, Holz, and Song]{exploitgym}
Zhun Wang, Nico Schiller, Hongwei Li, Srijiith~Sesha Narayana, Milad Nasr, Nicholas Carlini, Xiangyu Qi, Eric Wallace, Elie Bursztein, Luca Invernizzi, Kurt Thomas, Yan Shoshitaishvili, Wenbo Guo, Jingxuan He, Thorsten Holz, and Dawn Song.
\newblock Exploitgym: Can ai agents turn security vulnerabilities into real attacks?, 2026{\natexlab{a}}.
\newblock URL \url{https://arxiv.org/abs/2605.11086}.

\bibitem[Wang et~al.(2026{\natexlab{b}})Wang, Shi, He, Cai, Zhang, and Song]{wang2026cybergym}
Zhun Wang, Tianneng Shi, Jingxuan He, Matthew Cai, Jialin Zhang, and Dawn Song.
\newblock Cybergym: Evaluating {AI} agents' real-world cybersecurity capabilities at scale.
\newblock In \emph{The Fourteenth International Conference on Learning Representations}, 2026{\natexlab{b}}.
\newblock URL \url{https://openreview.net/forum?id=2YvbLQEdYt}.

\bibitem[Yang et~al.(2024)Yang, Jimenez, Wettig, Lieret, Yao, Narasimhan, and Press]{yang2024sweagent}
John Yang, Carlos~E. Jimenez, Alexander Wettig, Kilian Lieret, Shunyu Yao, Karthik Narasimhan, and Ofir Press.
\newblock {SWE}-agent: Agent-computer interfaces enable automated software engineering.
\newblock In \emph{Advances in Neural Information Processing Systems}, volume~37, 2024.
\newblock URL \url{https://papers.nips.cc/paper_files/paper/2024/hash/5a7c947568c1b1328ccc5230172e1e7c-Abstract-Conference.html}.

\bibitem[Yao et~al.(2023)Yao, Zhao, Yu, Du, Shafran, Narasimhan, and Cao]{yao2023react}
Shunyu Yao, Jeffrey Zhao, Dian Yu, Nan Du, Izhak Shafran, Karthik Narasimhan, and Yuan Cao.
\newblock {ReAct}: Synergizing reasoning and acting in language models.
\newblock In \emph{The Eleventh International Conference on Learning Representations}, 2023.
\newblock URL \url{https://openreview.net/forum?id=WE_vluYUL-X}.

\bibitem[Yu et~al.(2025)Yu, Chiang, Tsai, Huang, and Tsao]{primus2025}
Yao-Ching Yu, Tsun-Han Chiang, Cheng-Wei Tsai, Chien-Ming Huang, and Wen-Kwang Tsao.
\newblock Primus: A pioneering collection of open-source datasets for cybersecurity {LLM} training.
\newblock In \emph{Proceedings of the 2025 Conference on Empirical Methods in Natural Language Processing}, pp.\  10391--10413. Association for Computational Linguistics, 2025.
\newblock \doi{10.18653/v1/2025.emnlp-main.527}.
\newblock URL \url{https://aclanthology.org/2025.emnlp-main.527/}.

\bibitem[Zhang et~al.(2025)Zhang, Perry, Dulepet, Ji, Menders, Lin, Jones, Hussein, Liu, Jasper, Peetathawatchai, Glenn, Sivashankar, Zamoshchin, Glikbarg, Askaryar, Yang, Zhang, Alluri, Tran, Sangpisit, Yiorkadjis, Osele, Raghupathi, Boneh, Ho, and Liang]{zhang2025cybench}
Andy~K. Zhang, Neil Perry, Riya Dulepet, Joey Ji, Celeste Menders, Justin~W. Lin, Eliot Jones, Gashon Hussein, Samantha Liu, Donovan Jasper, Pura Peetathawatchai, Ari Glenn, Vikram Sivashankar, Daniel Zamoshchin, Leo Glikbarg, Derek Askaryar, Mike Yang, Teddy Zhang, Rishi Alluri, Nathan Tran, Rinnara Sangpisit, Polycarpos Yiorkadjis, Kenny Osele, Gautham Raghupathi, Dan Boneh, Daniel~E. Ho, and Percy Liang.
\newblock Cybench: A framework for evaluating cybersecurity capabilities and risks of language models.
\newblock In \emph{The Thirteenth International Conference on Learning Representations}, 2025.
\newblock URL \url{https://arxiv.org/abs/2408.08926}.

\bibitem[Zhou et~al.(2019)Zhou, Liu, Siow, Du, and Liu]{zhou2019devign}
Yaqin Zhou, Shangqing Liu, Jingkai Siow, Xiaoning Du, and Yang Liu.
\newblock Devign: Effective vulnerability identification by learning comprehensive program semantics via graph neural networks.
\newblock In \emph{Advances in Neural Information Processing Systems}, volume~32, 2019.
\newblock URL \url{https://papers.nips.cc/paper_files/paper/2019/hash/49265d2447bc3bbfe9e76306ce40a31f-Abstract.html}.

\bibitem[Zhu et~al.(2025)Zhu, Kellermann, Bowman, Li, Gupta, Danda, Fang, Jensen, Ihli, Benn, Geronimo, Dhir, Rao, Yu, Stone, and Kang]{zhu2025cvebench}
Yuxuan Zhu, Antony Kellermann, Dylan Bowman, Philip Li, Akul Gupta, Adarsh Danda, Richard Fang, Conner Jensen, Eric Ihli, Jason Benn, Jet Geronimo, Avi Dhir, Sudhit Rao, Kaicheng Yu, Twm Stone, and Daniel Kang.
\newblock {CVE-Bench}: A benchmark for {AI} agents' ability to exploit real-world web application vulnerabilities.
\newblock In \emph{Proceedings of the 42nd International Conference on Machine Learning}, volume 267 of \emph{Proceedings of Machine Learning Research}, pp.\  79850--79867, 2025.
\newblock URL \url{https://proceedings.mlr.press/v267/zhu25i.html}.

\bibitem[Zhuo et~al.(2026)Zhuo, Wang, Ding, Kumar, and Wang]{cyberzero}
Terry~Yue Zhuo, Dingmin Wang, Hantian Ding, Varun Kumar, and Zijian Wang.
\newblock Cyber-zero: Training cybersecurity agents without runtime.
\newblock In \emph{The Fourteenth International Conference on Learning Representations}, 2026.
\newblock \doi{10.48550/arXiv.2508.00910}.
\newblock URL \url{https://proceedings.iclr.cc/paper_files/paper/2026/file/f9f54762cbb4fe4dbffdd4f792c31221-Paper-Conference.pdf}.

\end{thebibliography}
